\documentclass[preprint,aps,prb,showpacs]{revtex4}
\usepackage{bm}
\usepackage{graphicx}
\usepackage{amsmath}
\usepackage{eufrak}
\usepackage{color}
\newcommand{\nix}[1]{}

\begin{document}

\title{All-electric detection of the polarization state of terahertz laser radiation}

\author{S.D.~Ganichev,$^{1}$\footnote{e-mail: sergey.ganichev@physik.uni-regensburg.de}
W.~Weber,$^1$ J.~Kiermaier,$^1$ S.N.~Danilov,$^1$ P.~Olbrich,$^1$
D.~Schuh,$^1$  W.~Wegscheider,$^1$ D.~Bougeard,$^2$ G.~Abstreiter,$^2$
and W.~Prettl$^1$}
\affiliation{$^1$  Terahertz Center, University of Regensburg, 93040,
Regensburg, Germany\\
$^2$  Walter Schottky Institute, TU Munich, Garching, Germany}

\begin{abstract}
Two types of room temperature detectors of terahertz laser
radiation have been developed which allow in an all-electric
manner to determine the plane of polarization of linearly
polarized radiation and the Stokes parameters of elliptically polarized
radiation,  respectively. The operation of the detectors is based
on photogalvanic effects in semiconductor quantum well structures
of low symmetry. The photogalvanic effects have nanoseconds
time constants at room temperature making  a high time resolution
of the polarization detectors possible.
\end{abstract}


\maketitle




\section{Introduction}

Terahertz (THz) physics and technology are on the frontiers of physics
holding great promise for progress in various  fields of science such as solid state
physics, astrophysics, plasma physics and others (see e.g.~\cite{sakai05,book,miles07,woolard07}).
Furthermore, THz physics
presents a potential for  applications  in medicine, environmental monitoring,
high-speed communication, security, spectroscopy of different  materials, including
explosives, etc.~\cite{miles07,woolard07,edwards00,shur03} Areas of THz physics of high current interest are the
development and application of coherent  semiconductor  sources, molecular gas lasers,
ultrafast time-domain spectroscopy based on femtosecond
near-infrared radiation pulses,  as well as the development of
novel detectors of laser radiation. An important characteristic of THz laser radiation is
its  state of polarization. The detection of the polarization state,
in particular the orientation of the electric field vector of linear
polarized radiation and/or the ellipticity  of transmitted, reflected or scattered, light
represents a powerful technique for analyzing the optical
anisotropy of various media such as solids, solid surfaces, plasmas,
and biological tissues.
The established way to gain information
about the polarization state is the use of optical elements, which
allow to determine  optical path differences.

Here we report on an all-electric room temperature semiconductor detector systems
which provide  information about the polarization state of THz laser
radiation. The operation of  detectors is
based on the photogalvanic effects in  semiconductor quantum well
(QWs) structures of suitably low symmetry.
The time constant of  photogalvanic currents is determined by the
momentum relaxation time of electrons which is in the range of
picoseconds at room temperature. This allows to measure  the
ellipticity of THz laser radiation with sub-nanosecond time
resolution. Preliminary results demonstrating
the method have been published in ref.~\cite{APL2007}. Here we present a detailed study
of the detection principle comprising the phenomenological theory of the physical
effects used for detection, the coupling of detector signals to Stokes parameters~\cite{Stokes},
the application of the method to low intensity cw lasers, spectral
behaviour in the terahertz range, dependence on the angle of incidence and temperature.
Finally, we  extend the method to the detection of the azimuth angle of
linear polarized THz radiation.

\section{Principle of operation}

\subsection{Photogalvanic effect}

The photogalvanic effect denotes the generation of an electric
current in a homogeneous semiconductor sample by homogeneous
irradiation. This is in contrast to photovoltaic effects, like
used in solar cells,  where optically generated electric charges
are separated by potential barriers. Homogeneity is usually
realized in the terahertz range because of the weak  absorption of
radiation. On a macroscopic level the photogalvanic effect can be
described by writing the $dc$ current $\bm j$ in
powers of the Fourier amplitude of the electric field of radiation
$ \bm E (\omega)$ at the frequency $\omega$. The
first possibly
non-vanishing
term is given by~\cite{book,sturman,book2}
\begin{equation}
\label{currentgeneral11} j_{\lambda} = \sum_{\mu, \nu}
\chi_{\lambda \mu \nu} E_{\mu}(\omega) E^*_{\nu}(\omega) \:
\end{equation}
where the expansion
coefficient $\chi_{\lambda \mu \nu}$ is a third rank tensor and
$E^*_{\nu}(\omega ) = E_{\nu}(-\omega )$ is  the complex conjugate
of $E_{\nu}(\omega)$.

The external product $E_\mu E^*_\nu$ can be re-written as a sum of
a symmetric and an antisymmetric  product
\begin{equation}
\label{currentgeneral12}
E_\mu E^*_\nu = \{ E_\mu E^*_\nu\} + [E_\mu E^*_\nu] ,
\end{equation}
with $\{ E_\mu E^*_\nu\} = (E_\mu E^*_\nu + E_\nu E^*_\mu )/2$ and
$[E_\mu E^*_\nu ] = (E_\mu E^*_\nu - E_\nu E^*_\mu )/2$.
%
%
This decomposition  corresponds to a splitting of $E_\mu E^*_\nu$
into its real and  imaginary part. The symmetric term is real while
the antisymmetric term  is purely imaginary. Due to the contraction of
the tensor $\chi_{\lambda\mu \nu}$ with $E_\mu E^*_\nu$ the same
algebraic symmetries  are projected onto the last two indices of
$\chi_{\lambda\mu \nu}$. The real part of $\chi_{\lambda\mu \nu}$
is symmetric in the indices $\mu\nu$ whereas the imaginary part is
antisymmetric. Antisymmetric tensor index pairs can be reduced to
a single  pseudovector index using the Levi--Civita totally
antisymmetric tensor $\delta_{\rho\mu\nu}$. Applying this
simplification  we obtain for the current due to the antisymmetric
part of $E_\mu E^*_\nu$
\begin{eqnarray}
\label{currentgeneral14} \chi_{\lambda\mu\nu}[E_\mu E^*_\nu ] =
\sum_{\rho}
\gamma_{\lambda\rho}i (\bm E \times \bm E^*)_\rho 
=
\sum_{\rho}
\gamma_{\lambda\rho}\hat{e}_\rho P_{\rm circ}\:E_0^2,
\end{eqnarray}
%
where  $\gamma_{\lambda\rho}$ is a  real second rank pseudotensor,
$E_0 = |{\bm E}|$, $P_{\rm circ}$ and $\hat{\bm e} =
\bm q / q$  are the degree of light circular
polarization (helicity) and the unit vector pointing in the
direction of light propagation, respectively.

In summary we find for the total photocurrent
\begin{equation}
\label{currentgeneral15} j_{\lambda} =
\sum_{\mu, \nu}
\chi_{\lambda \mu \nu} \{E_{\mu} E^*_{\nu}\} +
\sum_{\rho}
\gamma_{\lambda \rho}\:\hat{e}_\rho P_{\rm circ}\:E_0^2 \:,
\end{equation}
where $\chi_{\lambda\mu \nu} = \chi_{\lambda \nu\mu}$. In this
equation  the photogalvanic effect is decomposed into two distinct
phenomena, the linear photogalvanic effect (LPGE) and the circular
photogalvanic effect (CPGE), described by the first and the second
term on the right-hand side, respectively~\cite{book,sturman,book2}.
We note that the second term  also describes  the optically induced spin-galvanic
effect~\cite{book,book2,Ganichev03p935,Nature02}. Both photogalvanic
currents  have been observed in various
semiconductors and are theoretically well understood (for reviews
see e.g.~\cite{book,sturman,book2,Ganichev03p935}
and references therein).

From Eq.~(\ref{currentgeneral15}) follows that photogalvanic currents are determined by
the degree of linear polarization and the orientation of the polarization
ellipse as well as the handedness of elliptical polarization. In the following
we  rewrite Eq.~(\ref{currentgeneral15}) for C$_s$ symmetry corresponding to the structure
applied for detection, first for linear polarized radiation (LPGE) and
afterwords for elliptically polarized radiation (both LPGE and CPGE).

\subsection{Linear polarization}

The LPGE makes possible to determine the orientation of
polarization of linearly polarized radiation. In fact LPGE
represents a microscopic ratchet.  The periodically alternating
electric field superimposes a directed motion on the thermal
velocity distribution of carriers in spite of the fact that the
oscillating field neither does  exert a net force on the carriers
nor induce a potential gradient. The directed motion is due to
non-symmetric random relaxation and scattering in the potential of
a non-centrosymmetric medium.
The linear photogalvanic effect is usually observed  under
linearly polarized optical excitation but may also occur under
elliptically polarized radiation. It is allowed only in
non-centrosymmetric media of piezoelectric crystal classes where
nonzero invariant components of the third rank tensor $\chi_{\lambda \mu
\nu}$ exist. LPGE was studied in bulk crystals and has  also been
observed in quantum wells.

The LPGE in bulk  crystals has been proposed
for detection of the plane of polarization of linearly polarized
radiation in ref.~\cite{Andrianov88}. In bulk GaAs
crystals of T$_d$ symmetry irradiation with linearly polarized radiation
propagating in the [111]-crystallographic direction
yields transverse currents along the [1$\bar{1}$0] and
[11$\bar{2}$] axes. After Eq.~(\ref{currentgeneral15}) these currents
are given by
\begin{equation}
\label{Td}
j_{[1 \bar{1} 0]} = C I\cdot \sin{2\alpha},\quad
j_{[1 1 \bar{2}]}  = C I\cdot \cos{2\alpha},
\end{equation}
where $\alpha$ is the angle between the  plane of polarization and
the [11$\bar{2}$] axis, $I$ is the radiation intensity,  and $C$ is a constant factor that both
currents have in common. Equations~(\ref{Td}) show that simultaneous
measurements of the two currents allows one immediately to
obtain $\alpha$, i.e. to determine the space orientation of the
radiation polarization plane.  A polarization analyzer made of
GaAs crystals has been proved to give a reasonable signal from
9\,$\mu$m to about 400\,$\mu$m wavelength at room temperature.

Application of GaAs quantum well structures extends the  material
class suitable for detection of linear polarization.
For the detection of the radiation polarization state the most convenient
geometry requires a normal incidence of radiation on the sample
surface. A symmetry analysis shows that in order to obtain an LPGE
photoresponse at normal incidence the symmetry of the QW structure
must be as low as  the point group C$_{s}$. This group contains
only two elements: the identity  and one mirror reflection. This
can easily be obtained in QW structures by choosing a
suitable crystallographic orientation. This condition is met, for
instance, in (113)- or asymmetric (110)-grown structures.
We introduce here a coordinate system $(xyz)$ defined by
\begin{equation}
x \parallel [1\bar{1}0],\quad y\parallel [33\bar{2}],\quad  z\parallel [113]
\label{coordinateB}
\end{equation}
which is convenient for (113)-grown samples used here for detectors.
In this coordinate system  $x$ is normal to the only non-identity symmetry
element of C$_s$, the mirror plane.

The point group C$_s$ allows  an LPGE current at normal incidence
of the radiation on the sample because in this case the tensor
{$\bm \chi$}  has  nonzero components $\chi_{xxy} =
\chi_{xyx}$, $\chi_{yxx}$ and $\chi_{y y y}$.
Then  after Eq.~(\ref{currentgeneral15})
the current is given by~\cite{book}
%
%

\begin{subequations}
\begin{eqnarray}
\label{currentequ42}
j_{x} &=& \chi_{xxy} \left( E_x E_y^* +
E_y E_x^* \right) \:, \\
\label{currentequ42b}
j_{y} &=& \chi_{yxx} |E_x|^2 + \chi_{y y y}
|E_y|^2  
\end{eqnarray}
\end{subequations}

yielding for linearly polarized light
\begin{subequations}
\begin{eqnarray}
\label{currentequ43}
j_{x} &=& \chi_{xxy}  \hat{e}_{z} E_0^2   \sin{2
\beta} \:, \\
\label{currentequ43b}
j_{y} &=& \left( \chi_{+} +
\chi_{-}  \cos{2 \beta} \right) \hat{e}_{z} E_0^2
\end{eqnarray}
\end{subequations}
where $\chi_{\pm} = (\chi_{y x x} \pm \chi_{y y y})/2$ and $\beta$ is
the azimuth angle between the plane of polarization defined by the
electric field vector and the $x$-coordinate.  A current response
due to LPGE is allowed for both $x$- and $y$-directions.
Like in the bulk detector
addressed above, Eqs.~(\ref{currentequ43}) and (\ref{currentequ43b}) show that measuring LPGE currents simultaneously in $x$ and
$y$ direction allows one to determine unambiguously the azimuth angle
$\beta$ of linearly polarized radiation.

\subsection{Elliptical polarization}

While LPGE gives an experimental access to the state of linear
polarization it can not be used to determine the ellipticity of
radiation. On the other hand, the circular photogalvanic effect
depends on the helicity  of the radiation field and cannot  be
induced by linearly polarized excitation. Hence, this effect  may
be applied to measure the ellipticity of radiation. The CPGE
occurs in gyrotropic media only, as it is mediated by a
second rank pseudotensor. On a macroscopic level the photocurrent
of CPGE is  described by the phenomenological
equation~(\ref{currentgeneral14}) yielding $j \propto E^2
P_{\rm circ}$, where
the radiation helicity  $P_{\rm circ}$  is
given by
\begin{equation}
\label{helicity}
P_{circ} = \frac{\left( | E_{\sigma +} |^2 - | E_{\sigma -}  |^2
\right)} { \left( | E_{\sigma +} |^2 + | E_{\sigma -} |^2
\right)},
\end{equation}
where $ \left| E_{\sigma +}  \right|$ and $\left| E_{\sigma -}
\right|$ are the amplitudes of right and left handed circularly
polarized radiation, respectively. The helicity can easily be varied
by passing linearly polarized light at normal incidence through a birefringent
$\lambda/4$-plate (see Figs.~\ref{setup1}~and \ref{phi1}).
In this case the rotation of the quarter-wave
plate  by the angle $\varphi$ between the optical axis  of the
$\lambda/4$-plate and the direction of the initial radiation
polarization changes  the azimuth angle, the shape of the
polarization ellipse and the orientation of the rotation of the
electric field vector (see Fig.~\ref{phi1}, top).

In structures of C$_s$ symmetry the photogalvanic current due to elliptically polarized
radiation at normal incidence is after Eq.~(\ref{currentgeneral15})
given by~\cite{book}
\begin{subequations}
\begin{eqnarray}
\label{currentequ44}
j_{x} &=&
\chi_{xxy}  \hat{e}_z  E_0^2 \sin{4 \varphi} + \gamma_{x z}  \hat{e}_z   E_0^2  P_{\rm
circ}
\sin{2 \varphi} \:, \\
\label{currentequ44b}
j_{y} &=& \left( \chi_{+} + \chi_{-}
\cos^2{2 \varphi} \right) \hat{e}_{z} E_0^2. 
\end{eqnarray}
\end{subequations}
%
The CPGE current is described by the the second term on
the right side of the first equation. It flows perpendicular to the
mirror reflection plane of C$_s$ corresponding to the $x$
coordinate being  parallel to $[1\bar{1}0]$  because  the tensor
$\bm \gamma$ has a non-vanishing component $\gamma_{x z}$.
The LPGE current, given by the  first  term on
the right side of the Eq.(\ref{currentequ44}) and by the Eq.(\ref{currentequ44b}),
can be generated in both $x$ and $y$ directions.
It reflects  only the projection of the electric field of the elliptical polarized
radiation on the $x$ and $y$ axes   and does not contain any
information about the radiation ellipticity.

Equation~(\ref{currentequ44})
shows that in general, the photogalvanic current excited
by elliptically polarized radiation consists
of CPGE and LPGE.  In fact the interference of both effects, CPGE and LPGE, have been
observed  in certain materials~\cite{Ganichev03p935}.
However, CPGE and LPGE  are completely independent phenomena. One or the other effect can
dominate the photocurrent. As we show below, by proper choice of materials, the
information about radiation helicity can be obtained by structures
with dominating contribution of the CPGE which is proportional to
the radiation helicity and carries the information about the
direction of polarization vector rotation. Additional detection of
a signal by a structure with dominating contribution of the LPGE
provides the necessary information about azimuth angle of the
polarization ellipse.

\subsection{Monitoring of power by photon drag effect}

For reference it is necessary to know the power of radiation which
must be determined by a polarization independent sensor. For this
purpose we apply the  photon drag effect that is based on the transfer
of linear momentum from photons to charge carriers in semiconductors.
Phenomenologically it is described by~\cite{book,book2}
\begin{eqnarray}
\label{pd}
j_l=\sum_{mno}T_{lmno}q_m E_n E_m^\star ,
\end{eqnarray}
where $\bm{T}$ is a forth rank tensor and $\bm{q}$ the
wavevector of radiation. As a detector element we used bulk n-Ge of T$_d$
symmetry and $\mathbf{q}\parallel [111]$ crystallographic direction picking
up the photon drag current along the same direction (see Fig.~\ref{setup1}).
In this configuration, taking the coordinate $z^\prime \parallel [111]$, we get
\begin{equation}
\label{pd2}
j_{z^\prime} = T q_{z^\prime} \left( |e_x^2| + |e_y^2|\right)  E_0^2
\end{equation}
where
\begin{equation}
T = T_{{z^\prime}{z^\prime}{x^\prime}{x^\prime}} =  T_{{z^\prime}{z^\prime}{y^\prime}{y^\prime}}
\end{equation}
in T$_d$ symmetry. The polarization term in Eq.~(\ref{pd2}) is equal one, $|e_x^2| + |e_y^2| = 1$, thus
the photocurrent is polarization independent.

\section{Experimental technique}

\subsection{Detector units}

\begin{figure}[t!]
\begin{center}
\includegraphics[width=8.25cm]{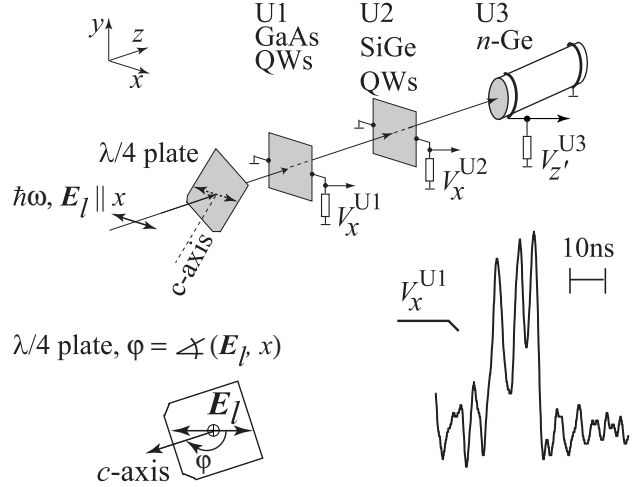}
\end{center}
\caption{Arrangement of the radiation ellipticity detector and experimental
set-up. The detector elements U1, U2, and U3 are GaAs QWs, a SiGe
QW structure,  and a $n$-Ge photon drag detector, respectively. The
signal voltages $V^{\rm U1}_x \propto j_x$,  $V^{\rm U2}_x \propto j_x$,  and
$V^{\rm U3}_{z'} \propto j_{z'}$   are picked up from the
elements U1, U2, and U3,
across load resistors of 50~$\Omega$, respectively. The ellipticity  of the radiation is varied by
passing linearly polarized laser radiation ($\bm{E}_l\parallel x$) through
a quarter-wave plate (bottom left). Bottom right shows  the
temporal structure of a typical signal pulse picked up by  the element U1  after
100 times voltage amplification in a bandwidth of
300~MHz and recorded by a  broad-band (1~GHz)
digital oscilloscope.}
\label{setup1}
\end{figure}

To realize these detector concepts we used three detector units,
U1, U2 and U3 (see Fig.~\ref{setup1}).
The U1 element is a (113)-oriented  MBE grown $p$-GaAs/Al$_{0.3}$GaAs$_{0.7}$
multiple QW structure containing 20 wells of 10~nm width with free
hole densities of about $2\cdot 10^{11}$~cm$^{-2}$ per QW. The U2
element, also (113)-oriented,  is an MBE grown
Si/Si$_{0.75}$Ge$_{0.25}$/Si single QW of 5~nm width.
The SiGe QW structure is one-side boron doped with a free carrier density in
the well of about $8 \cdot $10$^{11}$~cm$^{-2}$. For both square
shaped structures of 5x5~mm$^{2}$ size a pair of ohmic contacts is
centered on opposite sample edges along the $[1\bar{1}0]$
crystallographic axis.
The unit U2 has one additional pair of contacts
prepared along  $[33\bar{2}]$ crystallographic axis.
As the last element (U3)   we use a photon drag
detector for THz radiation~\cite{book,ch1Ganichev85pddet}. It consists
of a Ge:Sb cylinder of 5~mm diameter, which is  about 30~mm long. The crystal is grown along
$z^\prime \parallel [111]$-crystallographic direction. It has plane parallel end
faces and ring shaped electric contacts at both ends (see Fig.~\ref{setup1}).
The doping level is about $10^{14}$~cm$^{-3}$. The signal voltages are picked
up independently from each detector unit in a closed circuit
configuration across a 50~$\Omega$ load resistor. Signals are fed
into amplifiers with
voltage amplification by a factor of 100  and a bandwidth of
300~MHz and are measured by a digital  broad-band (1~GHz)
oscilloscope. For time resolved measurements the signal was picked up without
amplifier.

The functionality of the polarization detectors, their sensitivity
and time resolution are demonstrated using a pulsed  NH$_{3}$ THz
laser~\cite{book} with 100~ns pulses and a radiation power $P$ of
about 10~kW. Several lines of the NH$_{3}$ laser between $\lambda
= 76$~$\mu$m and 280~$\mu$m have been applied. We also use a $cw$
methanol laser with $P\approx$~20~mW in order to check the
detector applicability for detection of the 118.8~$\mu$m laser
line being important for tokamak applications (see
e.g.~\cite{tokomak1}). The excitation of our samples at room
temperature by  THz radiation results in  free carrier
(Drude-like) absorption~\cite{book}.

%
To demonstrate detection of ellipticity the  polarization of the laser beam has been modified from linear to elliptical
applying $\lambda/4$ plates  made of   $x$-cut crystalline quartz.  By that the
helicity of the incident light $P_{circ} =  \sin{2 \varphi}$ can be varied from
$P_{circ} =-1$ (left handed circular, $\sigma_-$) to $P_{circ}
=+1$ (right handed circular, $\sigma_+)$ by changing the angle
$\varphi$ between the initial linear polarization and
the optical axis (c-axis) of the quarter-wave quartz plate (see Fig.~\ref{setup1}).
In  Fig.~\ref{phi1}~(top)  the shape of
the polarization ellipse and the handedness of the radiation
are shown for various angles $\varphi$.

In experiment with linear polarization the plane of polarization has been
rotated applying  $\lambda/2$ plates also made of  $x$-cut
crystalline quartz, which enables us to vary the azimuth angle
$\beta$ between $0^\circ$ and $180^\circ$.
The rotation of the $\lambda/2$ plate by the angle $\beta/2$
between the initial linear polarization and the optical axis
of a half-wave quartz plate (c-axis)  leads to a
rotation of the polarization plane of linearly polarized radiation by
the angle $\beta$.

\subsection{Detector of elliptical polarization}

\begin{figure}[t!]
\begin{center}
\includegraphics[width=8.25cm]{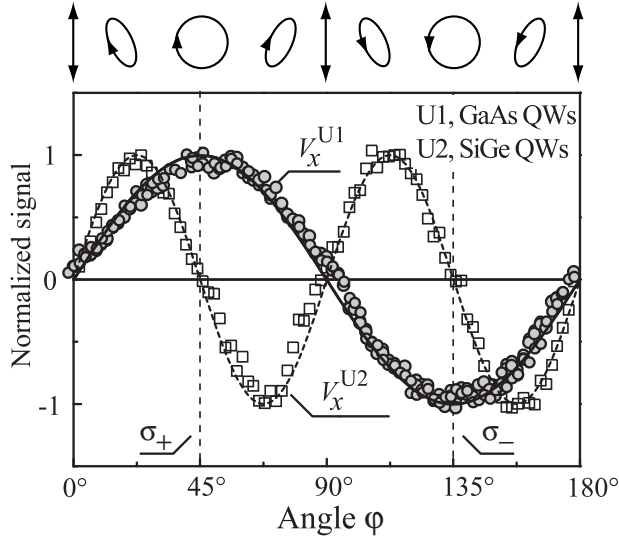}
\end{center}
\caption{Photoresponses $V^{\rm U1}_x$ of the GaAs QWs (U1) and  $V^{\rm U2}_x$
of SiGe QWs (U2)   as a function of the phase angle $\varphi$.
The signals are obtained at $\lambda = 148~\mu$m and at room temperature. The
maxima of the signal voltages are normalized to one. Lines
are fits after $V^{\rm U1}_x \propto sin 2\varphi$  (full line)  and $V^{\rm U2}_x \propto sin 4\varphi$ (dashed line)
for   elements U1 and  U2, respectively. These functional behaviour agrees with the polarization dependence of the photogalvanic current
given by the second  and  the first terms on the right hand side of Eq.~\protect(\ref{currentequ44}), correspondingly.
On top of the figure polarization ellipses  corresponding to various phase
angles $\varphi$ are plotted viewing from the direction toward which the wave approaching.}
\label{phi1}
\end{figure}

In response to irradiation  of the detector U1 made of the GaAs QW structure
we obtain   a voltage signal, $V_x^{\rm U1} \propto j_x$,
which changes its sign upon reversing
the helicity. Figure~\ref{phi1}  shows the photocurrent as a
function of the phase angle $\varphi$  revealing that  the
signal of the detector unit U1 closely follows the radiation helicity
($j_x \propto P_{\rm circ} = \sin 2 \varphi$).
%
In the case of the linearly polarized radiation, corresponding to $\varphi = 0^{\circ}$ or
$90^{\circ}$, the signal $V_x^{\rm U1}$ vanishes.

\begin{figure}[t!]
\begin{center}
\includegraphics[width=8.25cm]{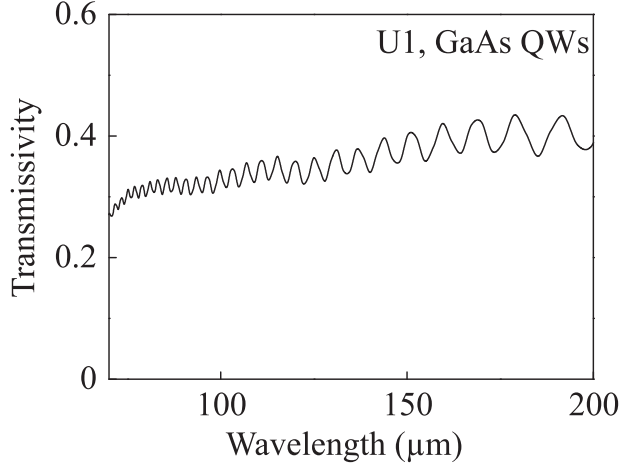}
\end{center}
\caption{Transmissivity of detector element U1 obtained by FTIR spectroscopy. }
\label{transmission}
\end{figure}

Passing the unit U1 radiation hits U2. This is possible because U1 is practically transparent
in the whole THz range.
This is demonstrated by FTIR measurements. Figure~\ref{transmission} shows the
transmissivity of the unit U1 obtained
in the spectral range from 70~$\mu$m to 200~$\mu$m. The
transmissivity in the whole range is
about 40 percent which just corresponds to the reflectivity of the sample made of GaAs.
Periodical modulation of the  spectrum is due to  interference
in the plane-parallel semiconductor structure.
The magnitude of the reflection and interference effects can be reduced by
anti-reflection coatings improving the sensitivity of the detector system.
Nearly the same transmission spectrum has been measured for the unit U2.
Low losses in detector units allows one to stack them one behind the other for illumination with the same
laser beam.

Irradiation of the  the second detector unit U2 made of the SiGe QW structure
also results in a signal $V_x^{\rm U2} \propto j_x$ depending on $\varphi$ which we picked-up from pair of contacts along $x$.
In contrast to U1,  the signal detected by U2 vanishes for circularly polarized radiation
and is given by $j_x \propto \sin 4\varphi$ as depicted in  Fig.~\ref{phi1}.

\begin{figure}[t!]
\begin{center}
\includegraphics[width=14.25cm]{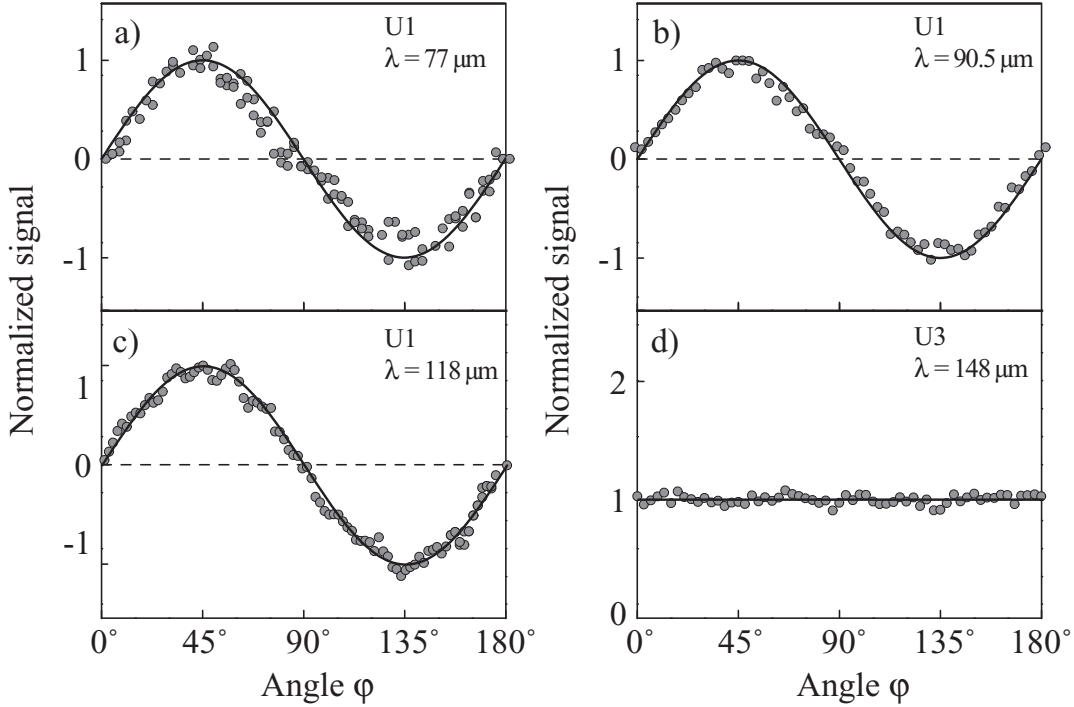}
\end{center}
\caption{Photoresponses as a function of the phase angle $\varphi$
obtained for various wavelengths.  The maxima of the signal voltages
are normalized to one.
(a)-(c): photoresponse $V^{\rm U1}_x$ of the GaAs QWs (U1). We emphasize that  the
signal  plotted in (c) is obtained by a cw
laser with about 20~mW power. Lines
are fits after $V^{\rm U1}_x \propto sin 2\varphi$.
d) Photoresponse   $V^{\rm U3}_{z'}$ of photon drag detector
element U3 demonstrating its  independence of polarization state.
Line is a fit after \protect Eq.~\ref{pd2}.
}
\label{phi2lambda}
\end{figure}

The observed  angular dependencies, $V_x^{\rm U1} \propto \sin 2 \varphi$ in the element U1 and
$V_x^{\rm U2} \propto \sin 4 \varphi $ in the element U2,  demonstrate that the
photocurrent is caused predominately by CPGE (Eq.~(\ref{currentequ44}),
second term on the right hand side) in our GaAs QWs, and
by LPGE (Eq.~(\ref{currentequ44}), first term on the right hand side) in the
present SiGe structures. We note that the dominance of one or the other effect (CPGE or LPGE)
in a material is not a matter of course. In other samples of the same crystallographic orientation
both may contribute with comparable strength yielding a beating in the $\varphi$
dependence of the photocurrent~\cite{book}.
The last detector unit U3 is needed to monitor the radiation power.
This unit is based on the photon drag effect whose sensitivity in the present geometry
is independent on the polarization state of radiation as shown in Fig.~\ref{phi2lambda}d.

From Figs.~\ref{phi1} and \ref{phi2lambda}d  follows that simultaneous
measurements of all three signals allows the unambiguous
determination of the radiation ellipticity and consequently the Stokes parameters~\cite{Stokes}.
Indeed, a  pair of signals of U1 and U2 obtained at an angle $\varphi$ is unique and
this pair of voltages will never be repeated for variation of
$\varphi$ between zero and $\pi$. The ratio of the signals of U1
and U2 - and even the sign of this ratio - is different for
different angles $\varphi$. The angles $\varphi$ can be easily determined from
measured voltages according  to
\begin{equation}
\varphi =  \frac{1}{2} \arccos{\left( \frac{V_x^{\rm U1}}{V_x^{\rm U2}} \right)}.
\end{equation}
Combining these measurements with the signal from U3, which yields the radiation power
$V_x^{\rm U3} \propto P \propto E_0^2$,
we get straight forward the Stokes parameters which completely characterize the state of polarization
of the radiation field. The Stokes parameters $s_0$ to $s_3$ are related to our measured
quantities $\varphi$ and $E_0^2$ by
\[s_0 = E_0^2\]
\[s_1 = s_0 \frac{1+\cos{4\varphi}}{2}\]
\[s_2 = s_0 \frac{\sin{4\varphi}}{2}\]
\[s_3 = -s_0 \sin{2\varphi}\]

%

Sensitivities of the detector units U1, U2 and U3 at the
wavelength of 148~$\mu $m are 3.2~mV/kW ($\varphi = 45^\circ$),
1.2~mV/kW ($\varphi = 22.5^\circ$), and 35~mV/kW, respectively.
These values are obtained with 50~$\Omega $ load resistors
and 100 times voltage amplification for  angles $\varphi$ corresponding to the
maximum of the voltage signal. Fig.~\ref{phi2lambda} shows the $\varphi$ dependence of the signal detected by the unit U1  for three different wavelengths ranging
from 77~$\mu$m to 118~$\mu$m.
The figure shows that the CPGE dominates the signal in this
spectral range and thus  gives an access to the handiness of radiation.
%
%
We emphasize that the $\lambda = 118$~$\mu$m data (Fig.~\ref{phi2lambda}c) are obtained with a cw optically pumped laser at a power of not more than $P \approx 20$~mW.
For this measurements we modulated our beam by a chopper with modulation frequency 353~Hz and used a low-noise pre-amplifier
(100 times voltage amplification) and a lock-in-amplifier for signal recording. We note that in this low modulation frequency
case the high time resolution of the set-up is not required. Thus we applied an open-circuit configuration
connecting U1 directly to the pre-amplifier with an  input impedance of  about 100 M$\Omega$
which resulted in the increase of the output voltage by about 30 times.
Generally, if the sub-nanosecond time resolution is not needed,
the output voltage at fixed radiation intensity can be substantially increased by more than one order of  magnitude
by increasing of the load resistance.
Variation of operation temperature by $\pm 30$~K
of both U1 and U2 units  did not show any considerable change of the sensitivity.
As a  large dynamic range is important for detection of laser radiation, we investigated
the dependence of the sensitivity of the detection system on the radiation intensity applying
$cw$ and high-power pulsed radiation. We observed that the ellipticity detector remains
linear up to 2~MW/cm$^{2}$ over more than nine orders of magnitude. This is indeed not
surprising because all detector units work on effects caused by Drude absorption at room temperature.

\begin{figure}[t!]
\begin{center}
\includegraphics[width=8.25cm]{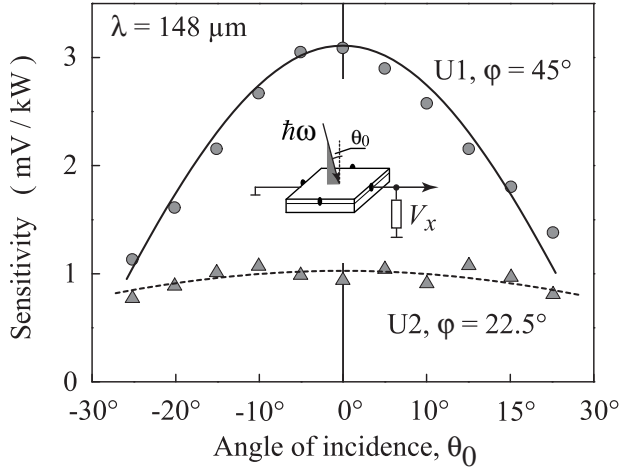}
\end{center}
\caption{Sensitivity of detector elements U1 and U2
as a function of the  angle of incidence $\theta_0$
obtained for $\varphi = 45^\circ $ and 22.5$^\circ$, respectively.
Dashed line shows the fit of the data obtained by the detector unit U2
after Eq.~(\protect \ref{projection}). Full line shows the fit of the data obtained by the detector unit U1
after the same Eq.~(\protect \ref{projection}), but taken into account an additional contribution
of the circular photon drag effect given by Eq.~(\protect \ref{CPDE}).
}

\label{theta}
\end{figure}

In a further experiment we checked the variation of the sensitivity due to a
deviation from normal incidence of  radiation.
The angle of incidence dependence of the signal for a fixed radiation helicity
($\varphi = 45^\circ$ for U1 and $\varphi = 22.5^\circ$ for U2) are  shown in
Fig.~\ref{theta} in comparison to calculations of photogalvanic currents
in QW structures of C$_s$ symmetry relevant to the present experiment.
As it  follows from Eq.~(\ref{currentequ44})
the dependence of the
CPGE and LPGE currents on the angle of incidence $\theta_0$
is determined by the value of the projection $\hat{\bm e}$ on the $z$-
axis and given  by~\cite{book}
\begin{equation}
\label{projection}
j_x \propto \hat{e}_{z} =  t_p t_s \cos \theta,
\end{equation}
where $\theta$ is the refraction angle defined by $\sin{\theta} =
\sin{\theta_0}/n_\omega$   and  the product of the transmission
coefficients $t_p$ and $t_s$ for linear $p$ and $s$ polarizations,
respectively,  after Fresnel's formula is given by
\begin{equation}
t_p t_s= \frac{4 n_\omega \cos^2{\theta_0}} {\Bigl(\cos{\theta_0} +
\sqrt{n_\omega^2- \sin^2{\theta_0}}\Bigr) \Bigl(n_\omega^2
\cos{\theta_0} + \sqrt{n_\omega^2- \sin^2{\theta_0}} \Bigr)}\:\:.
\label{Ch7currentequ18}
\end{equation}

While the angle of incidence dependence of the unit U2 agrees fully to Eq.~(\ref{projection}) (see dashed line in Fig.~\ref{theta}),
it does not sufficiently describe the signal of unit U1 at large angles of incidence.
However, taking into account the circular \textit{photon drag} effect~\cite{CPDE} in unit U1 a
good agreement is obtained (see solid line in Fig.~\ref{theta}).
The circular photon drag effect is like CPGE proportional to the $P_{circ}$
and hence does not affect the basic principle of operation.
Its dependence on the angle of incidence is given by~\cite{CPDE}
\begin{equation}
\label{CPDE}
j_x \propto t_p t_s \sin^2 \theta  P_{\rm circ}.
\end{equation}
Thus the presence of the current contribution due to the circular photon drag effect
only diminishes the value of the total photocurrent at large angles of incidence.
As a result we find that our whole detection system shows a tolerance
of  -10$^{\circ}$ to +10$^{\circ}$ for the angle of incidence.

The response time of each detector units is  due to free carrier
momentum relaxation and is in the order of ten picoseconds at room
temperature. The real time resolution, however, is RC-limited by
the design of the electric circuitry and by the bandwidth of
cables and amplifiers. A typical signal pulse of the unit U1
recorded by a broad-band (1~GHz) digital oscilloscope after 100
times voltage amplification in a bandwidth of 300~MHz is shown in
Fig.~\ref{setup1}. Additionally we performed measurements on the
unit U1 making use of the short pulse duration of the free
electron laser ''FELIX'' at Rijnhuizen in The
Netherlands~\cite{Knippels99p1578}. The FELIX operated in the
spectral range between 70~$\mu$m and 120~$\mu$m. The output pulses
of light from FELIX were chosen to be 3 ps long, separated by
40~ns, in a train (or ''macropulse'') of 5~$\mu$s duration. The
macropulses had a repetition rate of 5~Hz. In response to 3~ps
pulses we observed that the response time of U1 is determined by
the time resolution of our set-up,  but it is at least 100~ps or
shorter. This fast response is typical for photogalvanics where
signal decay time is expected to be of the order of the momentum
relaxation time\cite{book,sturman,book2} being in our samples at
room temperature of the order of  ten picoseconds.

\subsection{Detector of linear polarization}

\begin{figure}[t!]
\begin{center}
\includegraphics[width=8.25cm]{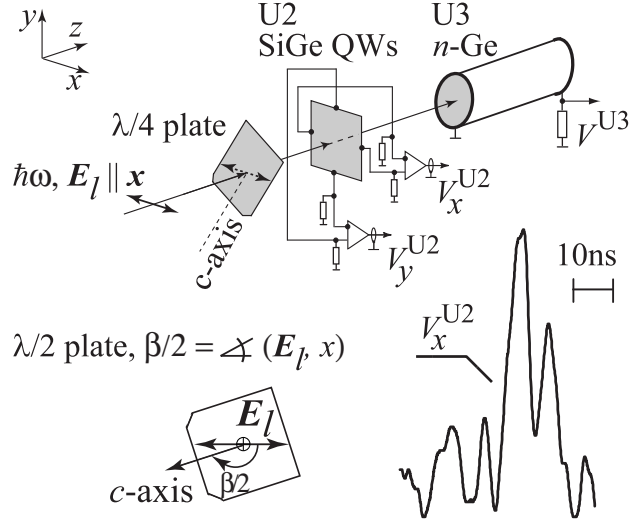}
\end{center}
\caption{Detector for the polarization of linearly
polarized radiation and experimental arrangement. The detector
elements U2 and U3 are a SiGe QW structure and an n-Ge photon drag
detector, respectively. The load resistors are
50~$\Omega$. The signals are fed into differential
amplifiers because a common  ground must be avoided and all contacts must be floating.
The orientation of the plane of polarization is
varied by passing  linearly polarized radiation ($\bm{E}_l\parallel
x$) through  a half-wave plate. The inset shows the
temporal structure of a typical signal pulse of the element U2
after 100 times voltage amplfication in a bandwidth of
300~MHz and recorded by a broad-band (1~GHz)
digital oscilloscope.}
\label{setup2}
\end{figure}

The  detector of linear polarization is sketched in
Fig.~\ref{setup2}. The arrangement is indeed the same as that of the
ellipticity detector (see Fig.~\ref{setup1})
with the only difference that unit U1 is not used and at U2 the current  signals
of both contact pairs are picked up simultaneously. We note that U1 can stay
in place because it is transparent.

The photocurrents  $j_x$ and $j_y$ of U2
are measured by the voltage drops $V_x^{\rm U2}$ and $V_y^{\rm U2}$ across
50~$\Omega$ load resistors in closed circuits. The signals are fed into differential
amplifiers
because a common  electric ground must be avoided and all electric potentials must be floating
(see Fig.~\ref{setup2}).

\begin{figure}[t!]
\begin{center}
\includegraphics[width=8.25cm]{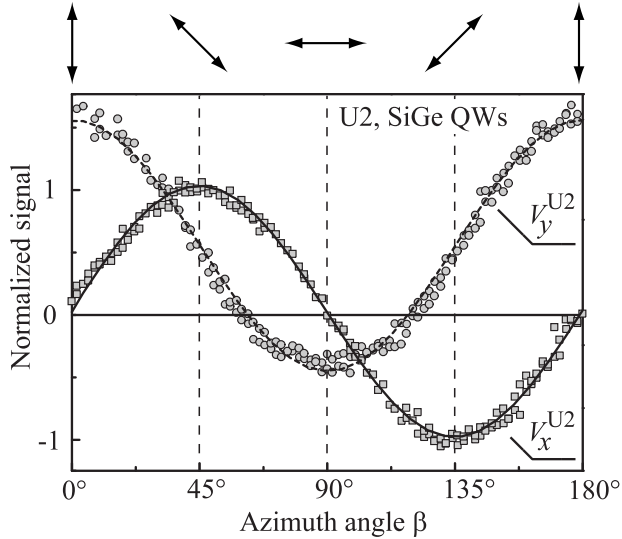}
\end{center}
\caption{Photoresponses $V^{\rm U2}_x$  and  $V^{\rm U2}_y$
of the unit U2 (SiGe QWs). The signal voltages $V^{\rm U2}_1$, and $V^{\rm U2}_2$ are
picked up by two differential amplifiers and are consequently proportional to
the currents $j_x$ and $j_y$.
The signals are obtained at $\lambda = 148~\mu$m and at room temperature. The
maxima of the signal voltages are normalized to one. Full and dashed lines
are fits after azimuth angle dependences given by  Eqs.~\protect(\ref{currentequ43}) and (\ref{currentequ43b}).
On top the polarization direction corresponding to various azimuth angles $\beta$ are
plotted viewing from the direction toward which the wave approaching.}
\label{beta}
\end{figure}

We note that in both amplifiers one of the inputs should be the inverted input line. These inputs are  marked in Fig.~\ref{setup2} by dots.
Figure~\ref{beta} shows a measurement of both signals as a function of
the polarization angle $\beta$. The full lines are fits after azimuth angle dependences given by
Eqs.~(\ref{currentequ43}) and (\ref{currentequ43b}) demonstrating an excellent agreement to the
experimental data. At the wavelength of 148~$\mu$m the sensitivities of the detector
unit U2  obtained with 100 times voltage amplification are 8~mV/kW
and 11~mV/kW for $\beta =
135^\circ$ and $\beta= 0^\circ$ for the signals $V_x^{\rm U2}$ and $V_y^{\rm U2}$,
respectively. In each case the  angle $\beta$ was adjusted to maximize  the signal.
Irradiating the detector by linear polarized light with  unknown
polarization angle $\beta$ yields two signals, $V_x^{\rm U2}$ and $V_y^{\rm U2}$,
whose ratio unambiguously gives  the value of the azimuth angle $\beta$. This angle
is obtained by solving the equation system
Eqs.~(\ref{currentequ43}) and (\ref{currentequ43b}). The polarization independent unit U3 is again used to monitor the power.
We checked that its sensitivity is, as in the case of elliptical polarization, 35~mV/kW
at the wavelength of 148~$\mu$m and it is independent on the angle $\beta$.

Our measurements demonstrate that all features of U2 in response to linear polarized radiation,
such as spectral characteristic and time resolution, tolerance in angle of incidence and dynamic
range are exactly the same as those in response to elliptically polarized radiation described above.
This is because in both cases the basic mechanism of response is the LPGE as
given by the first term of the right hand side in Eq.~(\ref{currentgeneral15}). This term  takes the form of
Eqs.~(\ref{currentequ43}) and~(\ref{currentequ43b}) for linear polarized radiation as well as
the form of the first term on the right hand side of Eq.~(\ref{currentequ44}) and
Eq.~(\ref{currentequ44b}) for elliptically polarized radiation.
In fact the tensor $\bm\chi$ describing all physical characteristics remains the same and
only the projection of the electric radiation field on a crystallographic direction
determines the polarization dependence of the current.

\section{Outlook}

The sensitivities of the detector systems presented here are
sufficient to detect short THz pulses of sources like optically
pumped molecular lasers and free electron lasers. The method has
also been successfully applied to monitor the ellipticity of cw
THz laser radiation by reducing the bandwidth of detection.
However, we would like to point out that the sensitivity  can be
improved essentially by using a larger number of QWs. Further
increase of sensitivity can be obtained by applying narrow band
materials such as HgTe QWs.  Most recently  we  observed in HgTe
QWs CPGE signals more than one order of magnitude larger than that
of the GaAs QWs investigated here~\cite{HgTe}. Another way to
improve sensitivity might be the application of specially designed
lateral structured QWs with enhanced asymmetry. Besides other
applications, sensitive THz detectors of the type shown here can
be of particular interest for the control of current density
profiles in plasmas which is important for tokamak
operation\cite{tokomak2,tokomak3,tokomak4,tokomak5,tokomak6}.
We also note that the CPGE and LPGE are also
observed at valence-to-conduction band
transitions~\cite{interband1,interband2,interband3}, at direct
intersubband transitions~\cite{book}, and in wide-band GaN semiconductor
heterojunctions~\cite{GaN}. Thus the applicability of the CPGE/LPGE
detection scheme may be well extended into the visible, the near infrared
and the mid-infrared spectral ranges.

\acknowledgments We thank L.E. Golub and E.L. Ivchenko for useful discussions and A.F.G. van der Meer
for assistance in doing of the time resolved measurements on FELIX.
The financial support from the DFG is gratefully acknowledged.


\end{document}